\documentclass[11pt,showpacs,floatfix]{revtex4-1}
\usepackage{graphicx,psfrag,amsmath,amssymb,amsfonts,latexsym,color,epsf,graphpap}

\begin{document}

\title{Ultraviolet cutoffs for quantum fields in cosmological spacetimes}
\author{Mauro El\'ias and Francisco D.~Mazzitelli}
\affiliation{Centro At\'omico Bariloche and Instituto Balseiro, 
Comisi\'on Nacional de Energ\'\i a At\'omica, 8400 Bariloche, Argentina.}

\date{today}

\begin{abstract} 
We analyze critically the renormalization of quantum fields in cosmological spacetimes, using
non covariant ultraviolet cutoffs. We compute explicitly the counterterms necessary to renormalize the 
semiclassical Einstein equations, using comoving and physical ultraviolet cutoffs. In the first case, the divergences 
renormalize bare conserved fluids, while in the second case it is necessary
to break the covariance of the bare theory. We point out that, in general,
the renormalized equations differ from those obtained with covariant methods, even after absorbing the infinities and choosing the 
renormalized parameters to force the consistency of the renormalized theory. 
We repeat the analysis for the evolution equation for the mean value of an interacting scalar field.

\end{abstract}
\maketitle
\section{Introduction}\label{sec:intro}
There is a well defined and rigorous approach for the renormalization of quantum fields in curved spacetimes \cite{books}. A covariant regularization of
 the theory (for example point splitting or dimensional regularization) is combined with the Hadamard structure of the two point function in order to perform 
adequate subtractions and obtain
 renormalized expressions for the effective action and the mean value of the Energy Momentum Tensor (EMT), the source in the so called Semiclassical Einstein 
 Equations (SEE). The infinities of the theory are absorbed into the bare constants of the classical gravitational action (Newton constant $G$,  cosmological 
 constant $\Lambda$, and dimensionless
 coefficients of terms quadratic in the curvature). The whole procedure preserves the covariance of the theory. 
 
 For different reasons, there have been attempts to regularize the theory using ultraviolet (UV) cutoffs. For example, in basic discussions of the cosmological 
 constant problem, 
 a three dimensional cutoff
 is considered in order to obtain a naive estimation of the zero point energy of quantum fields, which turns out to be an enormous contribution,
 and would require a
 fine tuning of the bare cosmological constant of 
 more than 120 orders of 
 magnitude \cite{reviewold}.  
  
 The use of UV cutoffs is more intuitive than dimensional regularization, but has been disregarded in semiclassical gravity because of its clash with the
 covariance of the theory \cite{fullingparker}. However, more recently it has been argued that it is possible, in principle, 
 to renormalize the theory in Friedmann-Lemaitre-Robertson-Walker (FLRW) spacetimes
 using a three dimensional {\it physical} cutoff \cite{Maggiore2012}. Being proportional to the scale factor, the time dependence of the physical cutoff 
 spoils the conservation of the EMT and, in order to restore it, it would be necessary to introduce non covariant counterterms, whose finite 
 parts should be carefully chosen to maintain the consistency of the renormalized theory. Some relevant aspects of the 
 proposal have not been worked out, like the calculation of the counterterms for general metrics, and a comparison with the usual approach.
 Regarding the quartic divergences that dominate the vacuum energy, and in order to avoid an unnatural fine tuning, the authors proposed \cite{Maggiore2012} a subtraction
 of the Minkowskian vacuum energy based on the Hamiltonian formulation of General Relativity.
 
 Alternatively, it has been suggested that the use of a {\it comoving} cutoff could provide a way to renormalize the theory without introducing non covariant
 counterterms \cite{Maroto1}. Once more, the explicit calculation of the counterterms have not been worked out, and in fact, as we will see,
 the particular subtraction used in that work is not compatible with a redefiniton of the bare constants of the theory. The comoving cutoff has also 
 been considered to compute 
 the effective potential for interacting fields in curved backgrounds \cite{Maroto2}. In these works it is assumed that
 the Fourier modes that effectively contribute to the vacuum energy have a maximum frequency much lower than the UV cutoff that sets the validity
 of the theory.
  
 The use of an UV physical cutoff has also been advocated by other authors. In Ref.\cite{Bilic} it was assumed that the non conservation of the EMT 
 is compensated by an additional non conserved source
 in Einstein equations. Alternatively, in Ref. \cite{Bernard}, only the $<T_{00}>$ component was computed using a cutoff, while the other non trivial components 
 $<T_{ii}>$ were determined from the SEE, in order to force the conservation of $<T_{\mu\nu}>$ and therefore the consistency of the SEE. In these works,
 there is no discussion about subtraction: on the contrary, while the quartic divergences are cancelled due to the supersymmetry of the theory \cite{Bilic},
 or assuming that vacuum fluctuations do not gravitate in Minkowski spacetime \cite{Bernard}, the quadratic 
 divergences are taken as the physical values of the EMT. UV cutoffs have also been considered when performing numerical calculations
 in the context of non-equilibrium field theory in cosmological spacetimes \cite{num}, and when computing loop corrections in inflationary models \cite{loops}.
 
  In this work we will not consider other interesting alternatives, like the use of a covariant cutoff based on the Schwinger proper 
 time approach \cite{shapiro}, or models in which the UV cutoff not only regulates the theory but also modifies
 the Lagrangian, giving rise to generalized dispersion relations for the quantum fields \cite{Kempf}.
    
The aim of this paper is to discuss in detail the renormalization procedure for quantum fields in cosmological spacetimes using as regulators comoving and physical three dimensional UV cutoffs.  Perhaps not sufficiently stressed in the recent previous works on the subject \cite{Maggiore2012, Maroto1, Maroto2}, in 
  addition to regularization, subtraction plays a crucial role in the renormalization process. Indeed,
 in order to absorb the infinities into the bare constants of the theory, the quantities to be subtracted should be generic, i.e., should not depend
on the particular metric considered. We will show that, using adiabatic subtraction \cite{books, fullingparker, Bunch} it is indeed possible to introduce counterterms and end up with a consistent theory.  However, the use of ultraviolet cutoffs make the choice of the counterterms rather unnatural. 
  We will also show that, while quartic and quadratic divergences are strongly dependent on the regularization method,   the logarithmic part is universal, and therefore the renormalization group flow of the gravitational constant is left unchanged. We will make a comparison of the resulting renormalized SEE, and show that they depend on the regularization scheme,
  and  differ from the renormalized SEE obtained using the standard covariant procedure.
 
 The paper is organized as follows. In Section II we describe the SEE in the context of FLRW spacetimes. In Section III we describe the renormalization of the SEE using comoving and physical cutoffs. We show that, for a comoving cutoff, it is necessary to introduce bare conserved fluids in order to absorb the divergences of the quantum EMT. For a physical cutoff, the divergences are a combination of non conserved tensors, and therefore it is necessary to include non covariant terms into the SEE to renormalize them.
We discuss in detail the relation with the usual renormalization approach, and make some comments related with the cosmological constant problem. In Section IV
 we consider a $\lambda\phi^4$ theory, and discuss the renormalization of the equation for the mean value of the field, once more using 
 non covariant methods. Section V contains the conclusions of our work.


\section{SEE in FLRW spacetimes}

In a curved spacetime with  a classical background metric $g_{\mu\nu}$, the SEE read
\begin{equation}
\label{RG1}
\frac{1}{8\pi G_B} G_{\mu \nu} + \Lambda_B g_{\mu \nu} + \alpha_{1B} H_{\mu \nu}^{(1)}+\alpha_{2B} H_{\mu \nu}^{(2)}+\alpha_{3B} H_{\mu \nu}= \langle T_{\mu \nu} \rangle + T^f_{B\mu \nu}, 
\end{equation}
where $G_{\mu\nu}$ is the Einstein tensor,  $H_{\mu\nu}^{(1,2)}$ and $H_{\mu\nu}$  are  tensors 
that come from the variation of quadratic terms in the curvature in the gravitational action ($R^2,R_{\mu\nu}R^{\mu\nu}$ and $R_{\mu\nu\rho\sigma}R^{\mu\nu\rho\sigma}$, respectively),  $\langle T_{\mu \nu} \rangle$ is the expectation value of the EMT of the quantum fields,  and  $T^f_{B\mu \nu}$ is the EMT of
a perfect fluid, that will be needed to renormalize the theory. The subindex $B$ indicates that the corresponding gravitational constants, even those contained
in $T_{B\mu\nu}^f$, are the bare ones.

The expectation value $\langle T_{\mu\nu} \rangle$ has UV divergences, that should be absorbed into the bare constants  of the theory. In order to make this point explicit,
we assume that $\langle T_{\mu\nu} \rangle$ is regularized in some way, and define its renormalized value as
\begin{equation}\label{sub}
\langle T_{\mu\nu} \rangle=\langle T_{\mu\nu} \rangle - \langle T_{\mu\nu} \rangle_{sub}+ \langle T_{\mu\nu} \rangle_{sub}\equiv   \langle T_{\mu\nu} \rangle_{ren}
+  \langle T_{\mu\nu} \rangle_{sub}\, ,
\end{equation}
where we have subtracted an appropriate tensor  $\langle T_{\mu\nu} \rangle_{sub}$ that cancels the divergences of $\langle T_{\mu \nu} \rangle$. Inserting 
Eq.(\ref{sub}) into Eq.(\ref{RG1}) we obtain
\begin{equation}
\label{RG11}
\frac{1}{8\pi G_B} G_{\mu \nu} + \Lambda_B g_{\mu \nu} + \alpha_{1B} H_{\mu \nu}^{(1)}+\alpha_{2B} H_{\mu \nu}^{(2)}+\alpha_{3B} H_{\mu \nu}= \langle T_{\mu \nu} \rangle_{ren} + \langle T_{\mu \nu} \rangle_{sub}+ T^f_{B\mu \nu}\, , 
\end{equation}
and, after absorbing the divergences of  $\langle T_{\mu \nu} \rangle_{sub}$ into the bare gravitational constants
\begin{equation}
\label{RG1R}
\frac{1}{8\pi G_R} G_{\mu \nu} + \Lambda_R g_{\mu \nu} + \alpha_{1R} H_{\mu \nu}^{(1)}+\alpha_{2R} H_{\mu \nu}^{(2)}+\alpha_{3R} H_{\mu \nu}= \langle T_{\mu \nu} \rangle_{ren} +\Delta\langle T_{\mu\nu}\rangle +  T^f_{R\mu \nu}\, .
\end{equation}
It is important to stress  that one should include in the theory all the bare terms
that are necessary to absorb the infinities of   $\langle T_{\mu \nu} \rangle_{sub}$.  Moreover, depending on the choice of $\langle T_{\mu \nu} \rangle_{sub}$
and of the regulator, it may happen that only a part of  $\langle T_{\mu \nu} \rangle_{sub}$ is absorbed into the bare terms. The remaining part would give an additional
contribution to the right hand side of the SEE, that we denoted by $\Delta\langle T_{\mu\nu}\rangle$ in Eq.(\ref{RG1R}).

In the rest of the paper we will consider spatially flat  FLRW background metrics,
\begin{equation}
ds^2=a(\tau)^2(d\tau^2-d\bar{x}^2)
\end{equation}where $\tau$ is the conformal time. For these metrics, and in four spacetime dimensions, the tensors  $H_{\mu\nu}^{(1,2)}$ and $H_{\mu\nu}$ are 
linearly dependent, and can be written in terms 
of   $H_{\mu\nu}^{(1)}$ \cite{books}. Therefore, we will set $\alpha_2=\alpha_3=0$ and denote $\alpha_1\equiv \alpha$. We stress that this is true only in four dimensions. 
Explicit expressions of the geometric tensors in FLRW metrics are presented in the Appendix.

The EMT of a classical fluid has energy density and pressure given by
\vspace{-1mm}
\begin{equation}
\label{tmunuf} 
 \rho =  T^f_{00}   / a^2  \quad \quad
 p =  T^f_{ii}   / a^2\, ,
\end{equation}
and the conservation equation reads
\begin{equation}
\rho ' + 3 \frac{a'}{a} (p+\rho) = 0 \label{conserv}
\end{equation}
where a prime denotes derivative with respect to $\tau$.

We will consider a free scalar field with classical action given by
\begin{equation}
S_{matter}= \int d^4x \sqrt{-g} \bigg(\frac{1}{2} \phi_{,\mu} \phi^{,\mu} - \frac{1}{2}(m^2+\xi R) \phi^2 \bigg)\, ,
\end{equation}
so the field equation reads
\vspace{-4mm}
\begin{equation}
(\Box + m^2 +\xi R) \phi = 0\, ,
\end{equation}
where $\xi$ is the coupling to the Ricci scalar curvature.  
Defining $\phi= \chi/a$, the Fourier modes of the field satisfy  
\begin{equation}\label{modes}
\chi''_k+\bigg[\omega_k^2+\bigg(\xi-\frac{1}{6}\bigg) a^2 R\bigg]\chi_k = 0\, ,
\end{equation}
where $\omega_k^2=k^2+m^2a^2$.

The EMT is given by \cite{books}
\begin{eqnarray}
\label{Tmunu}
T_{\mu\nu}=(1-2\xi)\partial_\mu \phi \partial_\nu \phi + \bigg(2\xi-\frac{1}{2}\bigg)g_{\mu\nu}\partial^\mu \phi \partial_\mu \phi - 2\xi \phi \nabla \partial_\nu \phi \nonumber
\\ + 2 \xi g_{\mu \nu} \phi \Box \phi - \xi G_{\mu \nu} \phi^2 + \frac{1}{2} m^2 g_{\mu \nu} \phi^2\, .
\end{eqnarray}

The vacuum expectation value of the energy-momentum tensor  can be written in terms of the Fourier modes of the field  $\chi_k$.
Taking into account the symmetries of the metric and vacuum state, the non-vanishing components are $\langle T_{00}\rangle$ and  $\langle T_{11} \rangle=\langle T_{22} \rangle = \langle T_{33} \rangle \equiv \langle T_{ii} \rangle$. Therefore $\langle T_{ii}\rangle=\frac{1}{3}(\langle T_{00}\rangle - a^2\langle T^\alpha_{\ \alpha}\rangle)$. 
The explicit expressions are
 \cite{Bunch} 
\begin{eqnarray}
\label{T00}\langle T_{00}\rangle &=&\frac{1}{4 \pi^2a^2}\int dk\,  k^2 \bigg[ |\chi'_k|^2+\omega^2_k |\chi_k|^2 + \bigg(\xi-\frac{1}{6}\bigg) \bigg(3D (\chi_k \chi_k^{*'}+\chi_k^{*}\chi_k') - \frac{3}{2}D^2 \chi_k^2 \bigg) \bigg]   \nonumber\\ 
&\equiv& \int dk \, \tilde T_{00}(k,\tau) \, ,
\nonumber\\  \langle T_{\ \alpha}^\alpha\rangle &=& \frac{1}{2 \pi^2 a^4}\int dk \, k^2 \bigg[ a^2 m^2  |\chi_k|^2 + 6 \bigg(\xi-\frac{1}{6}\bigg) \bigg(  |\chi'_k|^2 - \frac{a'}{a} (\chi_k \chi_k^{*'}+\chi_k^{*}\chi_k') - \omega^2_k |\chi_k|^2 \nonumber 
\\ && - \bigg( \frac{a''}{a} - \frac{a'^2}{a^2} \bigg) |\chi_k|^2 - 3\bigg(\xi-\frac{1}{6}\bigg) \bigg(D'+\frac{1}{2}D^2\bigg) |\chi_k|^2 \bigg) \bigg]\equiv \int dk\,   \tilde T_\alpha^\alpha(k,\tau)\, ,
\end{eqnarray}
where $D = 2 a'/a$.
The equations above have UV divergences coming from the high $k$ behavior of the Fourier modes.
In order to isolate the divergences, it is useful to consider the adiabatic approximation to the modes. Writing
\begin{equation}
\label{WKB}
\chi_k(\tau)=\frac{1}{\sqrt{2W_k(\tau)}} \exp{\bigg(-i\int^\tau W_k(\eta)d\eta}\bigg),
\end{equation}
Eq.(\ref{modes}) is equivalent to the nonlinear equation
\begin{equation}
W_k^2=\omega_k^2-\frac{1}{2}\left(\frac{ W''_k}{W_k}-\frac{3}{2}\frac{ {W'}_k^{2}}{W_k^2}\right)\, .
\label{nonl}
\end{equation}
In the large $k$ limit,  this equation can be solved iteratively using the number of derivatives of the scale factor as a small parameter. Doing this,  and 
inserting the solution into Eq.(\ref{T00}),  one obtains the adiabatic approximation to the EMT, $\langle T_{\mu\nu}\rangle_{ad}$, that 
can be written as the sum of its divergent and convergent parts
 \cite{Bunch}
\vspace{-5mm}
 \begin{equation}\label{Tdivcon}
 \langle T_{\mu\nu}\rangle_{ad}=\langle T_{\mu\nu}\rangle_{ad}^{div}+\langle T_{\mu\nu}\rangle_{ad}^{con}\, ,
 \end{equation}
\vspace{-3mm}
 where
\begin{eqnarray}
\label{TEM00}
\langle T_{00}\rangle_{ad}^{div} & = & \frac{1}{8\pi^2a^2} \int dk\ k^2\ \bigg[ 2\omega_k 
- \frac{3}{2} D^2\bigg(\xi-\frac{1}{6}\bigg)  \bigg( \frac{1}{\omega_k} + \frac{m^2a^2}{\omega_k^3} \bigg) \nonumber \\
&-& \frac{\big(\xi-\frac{1}{6}\big)^2}{16\omega_k^3}(72D''D-36D'^2-27D^4)  \bigg] \nonumber \\
\langle T_{\ \alpha}^{\alpha}\rangle_{ad}^{div} &=& \frac{1}{4\pi^2a^4} \int dk\ k^2 \bigg[ \frac{a^2m^2}{\omega_k} - \bigg(\xi-\frac{1}{6}\bigg) \bigg(\frac{3D'}{\omega}+\frac{a^2m^2}{\omega^3} \bigg(3D'+\frac{3}{4}D^2 \bigg) \bigg) \nonumber
\\& - & \frac{\big(\xi-\frac{1}{6}\big)^2}{4\omega_k^3} (18D'''-27D'D^2) \bigg]\, .
\end{eqnarray}
The explicit expressions for the convergent parts are written in the Appendix. Defining the adiabatic order as the number of derivatives of the scale factor, by  power counting it is easy to see that the zeroth adiabatic order diverges quartically, the second adiabatic order quadratically, and the fourth adiabatic order logarithmically. Higher adiabatic orders are finite. From these equations we see that the adiabatic expansion is a powerful tool to isolate the divergences, whatever the regularization used.  Moreover, as $\langle T_{\mu\nu}\rangle_{ad}$ can be written for any geometry in terms of time derivatives of the scale factor, it is the natural candidate
for  $\langle T_{\mu\nu}\rangle_{sub}$. 



\section{Cutoff regularization of the SEE}

A usual covariant approach to renormalize the theory in FLRW metrics is to combine dimensional regularization with adiabatic subtraction. Indeed, performing all calculations
in an arbitrary number of dimensions $n$, one can show that the divergences in $\langle T_{\mu\nu}\rangle_{ad}$  are all proportional to the geometric tensors  $g_{\mu\nu}, 
G_{\mu\nu}$, $H_{\mu\nu}$ and  $H_{\mu\nu}^{(1,2)}$. Therefore they can be absorbed into the gravitational constants $\Lambda_B, G_B,$ and $\alpha_{iB}$. A subtle point in the renormalization process is that the fourth adiabatic order contains  divergences that, as long as $n\neq 4$,  are proportional to geometric 
tensors and diverge as $1/(n-4)$, renormalizing the coupling constants $\alpha_{B1,2}$. However, 
in the limit $n\to 4$ they become  finite terms that modify $\langle T_{\mu\nu}\rangle_{ren}$, giving rise to the well known trace anomaly \cite{books}.

In this section we will explore the consequences of regularizing the theory imposing UV cutoffs in the momentum integrals. Our guiding principle will be that the divergences in 
 $\langle T_{\mu\nu}\rangle_{sub}$ should be absorbed into appropriate bare terms in the SEE, and this implies that the  subtraction should involve the adiabatic EMT up to the fourth adiabatic order. The last point needs further clarification. Using an UV cutoff, the renormalized EMT will be of the form
 \begin{equation}
 \langle T_{\mu\nu}\rangle_{ren} = \int_0^{k_{UV}}dk\,  \tilde T_{\mu\nu}(k,\tau)- \int_0^{k_{UV}}dk\,\tilde T_{\mu\nu}^{sub}(k,\tau)= \int_0^\infty dk\, ( \tilde T_{\mu\nu}(k,\tau)- \tilde T_{\mu\nu}^{sub}(k,\tau))\, ,
 \label{subk}
 \end{equation}
 where we made the reasonable assumption that $\langle T_{\mu\nu}\rangle_{sub}$ can be written as an integral over the momentum. The subtraction should cancel the
 UV divergences, and, at the same time,  should be written in terms of the scale factor and its derivatives. The last property is crucial in order to absorb the divergences into bare constants of the theory.  
 
 Note that when the subtraction is performed mode by mode,  as in the last equality in Eq.(\ref{subk}),  it is possible to take the limit $k_{UV}\to\infty$. 
 
 As in the high $k$ regime the modes of the field are well described by the WKB solutions to Eq.(\ref{modes}), the adiabatic
 modes will cancel the divergences. 
Therefore, the divergences of $\tilde T_{\mu\nu}$ will be cancelled by  $ \tilde T_{\mu\nu}^{ad}$, or at least by its divergent part. Moreover, in the adiabatic approximation the EMT becomes a local function of the scale factor and its derivatives, and this will allow for
 a redefinition of bare constants to cancel the divergences.  
 
 A simpler procedure would be to subtract the contribution of the modes with $k_{max}<k<k_{UV}$, as proposed in Ref.\cite{Maroto1}. In this case, the renormalized energy-momentum tensor would be
 \begin{equation}
 \langle T_{\mu\nu}\rangle_{ren} = \int_0^{k_{max}}dk \,\,  \tilde T_{\mu\nu}(k,\tau)\, . 
\end{equation}
While this is obviously a finite quantity, the divergent part should be computed for each metric in terms of the modes that solve Eq.(\ref{modes}), and the result would not be 
a local function of the metric and its derivatives. Therefore, the divergences could not be absorbed into bare constants of the theory.

\subsection{Comoving Cutoff}
\vspace{-3mm}
We will first consider  a comoving cutoff ($k_{UV}=\Lambda_c$). After evaluating  explicitly the integrals in Eq.(\ref{TEM00})  we obtain
\vspace{-5mm}
\begin{eqnarray}
\label{tmunucom}
\langle T_{00} \rangle_{ad}^{div}&=& \frac{\Lambda_c^4}{16\pi^2 a^2} + \frac{\Lambda_c^2 m^2}{16 \pi^2} - g_{00}\frac{m^4}{32\pi^2}\log[\Lambda_c/m] \nonumber 
\\&& - \frac{3 \Lambda_c^2 }{8\pi^2 a^2}\bigg(\xi-\frac{1}{6}\bigg) \frac{a'^2}{a^2} + \frac{m^2\big(\xi-\frac{1}{6}\big)}{8 \pi^2} G_{00} \log[\Lambda_c/m] \nonumber
\\ && + \frac{\big(\xi-\frac{1}{6}\big)^2}{16\pi^2} H^{(1)}_{00} \log[\Lambda_c/m]\nonumber
\\ [5mm]  \langle T_{ii} \rangle_{ad}^{div} &=& \frac{1}{3} \frac{\Lambda_c^4}{16\pi^2 a^2} - \frac{1}{3} \frac{\Lambda_c^2 m^2}{16 \pi^2} - g_{ii} \frac{m^4}{32\pi^2}\log[\Lambda_c/m] \nonumber
\\ && + \frac{\Lambda_c^2}{8\pi^2 a^2}\bigg(\xi-\frac{1}{6}\bigg)\bigg(2\frac{a''}{a} - 3 \frac{a'^2}{a^2} \bigg) + \frac{m^2 \big(\xi-\frac{1}{6}\big)}{8 \pi^2} G_{ii} \log[\Lambda_c/m] \nonumber
\\ && + \frac{\big(\xi-\frac{1}{6}\big)^2}{16 \pi^2} H_{ii}^{(1)} \log[\Lambda_c/m]\,,
\end{eqnarray}
where we omitted finite terms. 

One can readily check that the adiabatic EMT is a linear combination of conserved tensors, each one with a different dependence on $\Lambda_c$ and $m$.
For instance, the zeroth adiabatic order corresponds to the sum of three fluids:  a radiation fluid $(p=\rho/3)$ proportional to $\Lambda_c^4$,  a fluid with equation of state  $p=-\rho/3$ proportional to $m^2\Lambda_c^2$, and a cosmological constant ($p=-\rho)$ with a logarithmic divergence. 
The quadratic divergence of the second adiabatic order corresponds to a fluid with energy density proportional to $a'^2/a^6$. The logarithmic divergences
of all adiabatic orders are proportional to geometric tensors.

The main points of this calculation are that the introduction of a comoving UV cutoff does not spoil the conservation of the EMT, and that it requires the 
introduction of non-standard bare fluids into the SEE in order to absorb the divergences. In what follows will  compute explicitly the counterterms needed to 
renormalize the theory. The SEE with the appropriate bare fluid read
\vspace{-2mm}
\begin{equation}
\label{F111}
\frac{1}{8\pi G_B} \bigg( \frac{a'^2}{a^2} \bigg) + \Lambda_B g_{00} + \alpha_B H_{00}^{(1)} + \beta_B \frac{1}{a^2} + \gamma_B - \delta_B 3 \frac{a'^2}{a^4} = \langle T_{00} \rangle_{ren} + \langle T_{00} \rangle_{ad}
\end{equation}
\vspace{-16mm}
\begin{equation}
\label{F222}
\frac{1}{8\pi G_B} \bigg(2 \frac{a''}{a} - \frac{a'^2}{a^2} \bigg) + \Lambda_B g_{ii} + \alpha_B H_{ii}^{(1)} + \frac{\beta_B}{3} \frac{1}{a^2} - \frac{1}{3} \gamma_B + \delta_B \bigg( 2\frac{a''}{a^3} - 3\frac{a'^2}{a^4} \bigg) = \langle T_{ii} \rangle_{ren} + \langle T_{ii} \rangle_{ad}
\end{equation}
As anticipated, we introduced three new bare constants in the theory. The relations between bare and renormalized constants read, using minimal subtraction
\begin{eqnarray}
\Lambda_{ren} &=& \Lambda_B - \frac{m^4}{32\pi^2}\log[\Lambda_c/m]\nonumber
\\ \frac{1}{G_{ren}} &=& \frac{1}{G_B} - \frac{m^2\big(\xi-\frac{1}{6}\big)}{\pi}\log[\Lambda_c/m]\nonumber
\\ \alpha_{ren} &=& \alpha_B - \frac{\big(\xi-\frac{1}{6}\big)^2}{16\pi^2}\log [\Lambda_c/m]\nonumber
\\ \beta_{ren} &=& \beta_B - \frac{\Lambda_c^4}{16 \pi^2} \nonumber
\\ \gamma_{ren} &=& \gamma_B - \frac{m^2\Lambda_c^2}{16 \pi^2}\nonumber
\\ \delta_{ren} &=& \delta_B - \frac{\Lambda_c^2\big(\xi-\frac{1}{6}\big)}{8\pi^2}\, .
\end{eqnarray}
The SEE without additional classical fluids  can be obtained by setting $\beta_{ren}=0, \gamma_{ren}=0$ and $\delta_{ren}=0$. 

\vspace{-5mm}
\subsection{Physical Cutoff}

Instead of using a comoving cutoff, one could alternatively use a physical cutoff   $k_{UV}=a(\tau)\Lambda_{UV}$,  where $\Lambda_{UV}$ is constant.
The additional time dependence in the cutoff will compromise the conservation of the adiabatic EMT, and therefore the renormalization of the SEE
will involve non covariant counterterms. 

The divergent part of the adiabatic EMT is given by Eq.(\ref{tmunucom})  with the replacement $\Lambda_c\to a\Lambda_{UV}$
\begin{eqnarray}
\label{tmunuphys}
\langle T_{00} \rangle_{ad} &=& \bigg[ \frac{\Lambda_{UV}^4}{16\pi^2} + \frac{\Lambda_{UV}^2 m^2}{16 \pi^2} - \frac{m^4}{32\pi^2} \log[\Lambda_{UV}/m] \bigg] g_{00} \nonumber 
\\&+& \bigg[ \frac{\Lambda_{UV}^2 }{8\pi^2}\bigg(\xi-\frac{1}{6}\bigg)+ \frac{m^2\big(\xi-\frac{1}{6}\big)}{8 \pi^2} \log[\Lambda_{UV}/m] \bigg] G_{00} \nonumber
\\ & +& \frac{\big(\xi-\frac{1}{6}\big)^2}{16 \pi^2} H_{00}^{(1)} \log[\Lambda_{UV}/m]\nonumber
\\ \langle T_{ii} \rangle_{ad} &=& \bigg[- \frac{1}{3} \frac{\Lambda_{UV}^4}{16\pi^2} + \frac{1}{3} \frac{\Lambda_{UV}^2 m^2}{16 \pi^2} - \frac{m^4}{32\pi^2} \log[\Lambda_{UV}/m] \bigg] g_{ii} \nonumber
\\ & + & \bigg[ \frac{\Lambda_{UV}^2}{8\pi^2} \bigg(\xi-\frac{1}{6}\bigg) + \frac{m^2 \big(\xi-\frac{1}{6}\big)}{8 \pi^2} \log[\Lambda_{UV}/m] \bigg] G_{ii} - \frac{\Lambda_{UV}^2}{4\pi^2} \bigg(\xi-\frac{1}{6}\bigg) \frac{a'^2}{a^2} \nonumber
\\ & +& \frac{\big(\xi-\frac{1}{6}\big)^2}{16 \pi^2} H_{ii}^{(1)} \log[\Lambda_{UV}/m] \, ,
\end{eqnarray}
where again we omitted finite terms.

The above equations clearly show the non-conservation of the adiabatic EMT. The quartic divergence corresponds to a fluid with $p=1/3\rho=constant$, which is not conserved.
The same happens with the quadratic divergences of adiabatic orders zero and two. For example, the divergences  of adiabatic order two consist of a conserved term proportional to the Einstein tensor $G_{\mu\nu}$, plus a non conserved contribution proportional to $\Lambda_{UV}\delta_{\mu i}a'^2/a^2 $.  Note however that all logarithmic divergences are proportional to conserved 
geometric tensors.

In order to compute the counterterms, the SEE should be generalized introducing non-conserved bare terms (that can be thought as coming from the variation of 
non-covariant terms in the classical gravitational action \cite{Maggiore2012}). Thus, in the $ii$-semiclassical equation we split the Newton and cosmological constants in two different
constants, as follows
\begin{equation}
\label{F1}
\frac{1}{8\pi G_B} \bigg( \frac{a'^2}{a^2} \bigg) + \Lambda_B g_{00} + \alpha_B H_{00}^{(1)}= \langle T_{00} \rangle_{ren} + \langle T_{00} \rangle_{ad}
\end{equation}
\vspace{-5mm}
\begin{equation}
\label{F2}
\frac{1}{8\pi G_B} \bigg( 2 \frac{a''}{a} - \frac{a'^2}{a^2} \bigg) + \frac{1}{8\pi\tilde{G}_B}\frac{a'^2}{a^2} +  (\Lambda_B+\tilde \Lambda_B) g_{ii} + \alpha_B H_{ii}^{(1)}= \langle T_{ii} \rangle_{ren} + \langle T_{ii} \rangle_{ad}
\end{equation}
Inserting  Eq.(\ref{tmunuphys}) into Eqs.(\ref{F1}) and (\ref{F2}) we obtain
\begin{eqnarray}
\Lambda_{ren}&=& \Lambda_B - \frac{\Lambda_{UV}^4}{16\pi^2} - \frac{m^2 \Lambda_{UV}}{16\pi^2}+\frac{m^4}{32\pi^2}\log[\Lambda_{UV}/m]\nonumber
\\ \frac{1}{G_{ren}} &=& \frac{1}{G_B}-\frac{\Lambda_{UV}^2\big(\xi-\frac{1}{6}\big)}{\pi}-\frac{m^2\big(\xi-\frac{1}{6}\big)}{\pi}\log[\Lambda_{UV}/m]\nonumber
\\ \alpha_{ren} &=& \alpha_B - \frac{\big(\xi-\frac{1}{6}\big)^2}{16\pi^2}\log [\Lambda_{UV}/m]\nonumber\\
\tilde\Lambda_{ren}&=&\tilde \Lambda_B+\frac{4}{3} \frac{\Lambda_{UV}^4}{16\pi^2} + \frac{2}{3} \frac{m^2\Lambda_{UV}^2}{16\pi^2} \nonumber
\\ \frac{1}{\tilde{G}_{ren}} &=& \frac{1}{\tilde{G}_B} + \frac{\Lambda_{UV}^2\big(\xi-\frac{1}{6}\big)}{4\pi^2}\, .
\end{eqnarray}
The covariance of the SEE  is restored by setting to zero the renormalized values of the extra constants $1/\tilde G_{ren}=0$ and $\tilde\Lambda_{ren}=0$.

\subsection{Comparison with covariant regularization}
In the usual covariant approach (dimensional regularization plus adiabatic subtraction)
one subtracts
$\langle T_{\mu\nu}\rangle_{sub} =\langle T_{\mu\nu}\rangle_{ad}$, where the adiabatic EMT contains all  divergent and finite terms  up to the fourth adiabatic order. The calculations  should be entirely performed in $n$ dimensions from the beginning, and in this case it can be shown that both the finite and divergent terms of $\langle T_{\mu\nu}\rangle_{ad}$ can be 
absorbed into the gravitational constants of the theory \cite{foot1}. Moreover, when calculating $\langle T_{\mu\nu}\rangle _{ren}$ as the difference of two divergent integrals, the regulator can be removed performing the subtraction mode by mode.  One of the consequences of this procedure is the appearance of the trace anomaly, which is produced
by the finite terms in $\langle T_{\mu\nu}\rangle_{ad}$  subtracted from $\langle T_{\mu\nu}\rangle$. Indeed, the trace of the renormalized EMT reads, for
$m^2=0$ and $\xi=1/6$ \cite{books,Bunch}
\begin{equation}\label{tanom}
\langle T_\mu^\mu\rangle_{ren}=\frac{1}{960\pi^2a^4}(D'''-D'D^2)\, ,
\end{equation} while vanishes at the  classical level.

The situation in the presence of UV cutoff has some subtle points that deserve clarification. Up to now, we have not specified whether we subtracted the full adiabatic 
EMT or only its divergent part. In order to make contact with the covariant renormalization, we will subtract
the full adiabatic tensor \cite{foot2}.
 In this case, as the cutoff can be removed in the expression of $\langle T_{\mu\nu}\rangle_{ren}$ (see Eq.(\ref{subk})), all regularization methods,
covariant or not, give the same answer for $\langle T_{\mu\nu}\rangle_{ren}$. This does not mean, however, that the SEE are all equivalent.  Indeed, in the covariant approach, the {\it complete} $\langle T_{\mu\nu}\rangle_{ad}$ is absorbed into the gravitational bare constants. Therefore, $\langle T_{\mu\nu}\rangle_{ren}$  is the only quantum contribution to the
SEE.  When using UV cutoffs, only the divergent terms are cancelled by counterterms, and there are additional contributions coming from the finite parts of
$\langle T_{\mu\nu}\rangle_{ad}$, that we denoted by $\Delta\langle T_{\mu\nu}\rangle$ in Eq.(\ref{RG1R}).

For a comoving cutoff we have
\begin{eqnarray}
\label{extracom}
\Delta \langle T_{00}\rangle &=& \frac{m^4 a^2}{128\pi^2}(1+4\log(a/2)) -  \frac{\big(\xi-\frac{1}{6}\big)m^2}{16\pi^2} (3 + 2 \log(a/2)) G_{00} \nonumber \\
 &-& \frac{\big(\xi-\frac{1}{6}\big)^2}{16\pi^2} (1+\log(a/2)) H_{00}^{(1)} - \frac{m^2}{288 \pi^2}G_{00} - \frac{1}{15360 \pi^2 a^2}D^4
\nonumber\\
&+& \frac{1}{17280\pi^2} H_{00}^{(1)} - \frac{\big(\xi-\frac{1}{6}\big)}{288\pi^2} H_{00}^{(1)} + \frac{\big(\xi-\frac{1}{6}\big)^2}{64\pi^2a^2}(18D'D^2 + 9 D^4) \nonumber\\
\Delta \langle T_{ii}\rangle &=& -\frac{m^4 a^2}{384\pi^2}(7+12\log(a/2)) - \frac{m^2 \big(\xi-\frac{1}{6}\big)}{16\pi^2} (1+2\log(a/2)) G_{ii} \nonumber \\
&-& \frac{\big(\xi-\frac{1}{6}\big)^2}{16\pi^2} (1+\log(a/2)) H_{ii}^{(1)} - \frac{m^2}{288 \pi^2}G_{ii} - \frac{m^2(\xi-\frac{1}{6})}{8\pi^2} G_{ii} \nonumber \\
&-& \frac{m^2(\xi-\frac{1}{6})}{32\pi^2} D^2 + \frac{1}{5760 \pi^2a^2} \bigg(D'D^2-\frac{1}{8}D^4 \bigg) + \frac{1}{17280 \pi^2}H_{ii}^{(1)} \nonumber\\
&-& \frac{\big(\xi-\frac{1}{6}\big)}{288\pi^2}H_{ii}^{(1)} - \frac{\big(\xi-\frac{1}{6}\big)^2}{192\pi^2a^2}\bigg(72D''D+54D'^2+54D'D^2-\frac{45}{2}D^4\bigg) \, .
\end{eqnarray}
We stress that to compute  $\Delta \langle T_{\mu\nu}\rangle$ one should take into account not only the convergent part of the adiabatic EMT (see Appendix), but also the
finite terms coming from $\langle T_{\mu\nu}\rangle_{ad}^{div}$. 
One can check that $\Delta \langle T_{\mu\nu}\rangle$  is a conserved tensor, and that the contributions proportional 
to $\log(a)$ are crucial for the conservation. 

Unless the finite extra terms described in Eq.(\ref{extracom}) are artificially 
cancelled out  by new counterterms, the resulting SEE differ from the usual ones, since the quantum effects are described by the
effective EMT
$T_{\mu\nu}^{(eff)}=\langle T_{\mu\nu}\rangle_{ren}+\Delta \langle T_{\mu\nu}\rangle$.
 For example, in the conformal limit ($m^2=0$ and $\xi=1/6$),  this effective EMT is traceless. Indeed,  $\langle T_{\mu\nu}\rangle_{ren}$ 
 has the anomalous trace Eq.(\ref{tanom}),  which is exactly cancelled by $\Delta \langle T_\mu^\mu\rangle$. This can be readily checked by an
 explicit calculation, but it is to be expected because, while the full $\langle T_{\mu\nu}\rangle_{ad}$ is absorbed into the bare constants when using a 
 covariant regularization, this is not the case for an UV cutoff.
  
The analysis can be repeated using a physical cutoff. The expression for $\Delta \langle T_{\mu\nu}\rangle$ would be still given by Eq.(\ref{extracom}), omitting
 all terms proportional to $\log(a)$. Therefore,  the effective energy-momentum tensor would not conserved. For consistency of the SEE, the non conserved terms must 
 be cancelled by
 additional finite counterterms. The remaining conserved terms read
\begin{eqnarray}
\label{extrafis}
\Delta \langle T_{00}\rangle &=&  
- \frac{m^2}{288 \pi^2}G_{00} + \frac{1}{17280\pi^2} H_{00}^{(1)} - \frac{\big(\xi-\frac{1}{6}\big)}{288\pi^2}H_{00}^{(1)} -  \frac{1}{15360 \pi^2 a^2}D^4 \nonumber \\
\Delta \langle T_{ii}\rangle &=&
- \frac{m^2}{288 \pi^2 } G_{ii} + \frac{1}{17280\pi^2} H_{ii}^{(1)} - \frac{\big(\xi-\frac{1}{6}\big)}{288\pi^2} H_{ii}^{(0)} + \frac{1}{5760 \pi^2 a^2} \bigg(D'D^2-\frac{1}{8}D^4 \bigg)\, .
\end{eqnarray}
Note that the additional counterterms should cancel all terms proportional to $m^4, m^2(\xi-1/6)$ and $(\xi-1/6)^2$,  since they produce  non conserved contributions
to $\Delta\langle T_{\mu\nu}\rangle$ \cite{foot4}.  Note also that while the terms proportional to $G_{\mu\nu}$ and $H_{\mu\nu}^{(1)}$ can be absorbed 
into a finite redefinition of the gravitational constants, the remaining contributions to $\Delta \langle T_{\mu\nu}\rangle$ are nontrivial.
The effective EMT for a physical cutoff does not have a trace anomaly and differs from the one computed using a 
comoving cutoff by local terms.

\vspace{-5mm}

\subsection{A comment on the cosmological constant problem}
\vspace{-2mm}
The zero point energy of the quantum fields gives an enormous contribution to the cosmological constant \cite{reviewold}. A naive estimation in Minkowski spacetime consists in considering the sum of the ground state energy of each mode of the field. The sum is performed with a  three dimensional cutoff to be of the order of Planck mass. Taking this naive estimation as the value of the cosmological constant, it gives a disagreement of 122 orders of magnitude with respect to the observed value.
 However, as pointed out in Ref. \cite{akhmedov}, the three dimensional cutoff does not respect Lorentz invariance, and the problem  is alleviated when considering dimensional regularization, since this regularization kills the power law divergences,
and keep only the logarithmic ones.  Moreover, with  a three dimensional cutoff in Minkowski spacetime,  the vacuum expectation value of the EMT does not correspond
to a cosmological constant but to that of a radiation fluid. 

The results presented in this section show similar  characteristics, generalized to curved spacetimes. When using a comoving cutoff, the renormalization of the cosmological constant involves only a logarithmic divergence. However, it is necessary to introduce into the theory three new bare fluids, whose renormalized values should be fine tuned to zero.
The quartic divergences usually advocated as contributions to the cosmological constant, now give naive estimations to the amplitude of the bare fluids.
 On the other hand, when using a physical cutoff, in addition to the usual quartic contribution to both cosmological constants $\Lambda_R$ and $\tilde\Lambda_R$, it is necessary to fine tune  the renormalized value of the new constants ($\tilde\Lambda_R$ and $1/\tilde G_R$) to zero in order to respect the covariance of the theory at the renormalized level. 
      
These results reinforce the idea \cite{reviewccnew} that the power law divergences should not be taken seriously as estimations of the zero point energy contribution to the cosmological constant. After absorbing the infinities into the bare constants, a finite piece coming from the logarithmic divergence remains, and is of order
$m^4 \log[m/\mu]$, where $\mu$ is an arbitrary scale (this is still enormous compared with the observed value of the cosmological constant, even for the electron mass).

\section{Mean value equation for selfinteracting fields}

We will now consider a scalar field with selfinteraction $\lambda \phi^4$. The classical action is 
\begin{equation}
S_{matter}= \int d^4x \sqrt{-g} \bigg(\frac{1}{2} \phi_{,\mu} \phi^{,\mu} - \frac{1}{2}(m_B^2+\xi_B R) \phi^2 - \frac{\lambda_B}{4!} \phi^4\bigg)\, ,
\end{equation}
and the field equation
\vspace{-5mm}
\begin{equation}
\label{kgint}
\Box \phi + (m_B^2+\xi_B R) \phi + \frac{\lambda_B}{3!} \phi^3 = 0\, .
\end{equation}
We have written the action in terms of the bare constants of the theory.

We will be concerned with the renormalization of the equation for the mean value of the field $\phi_0=\langle\phi\rangle$. Writing  $\phi=\phi_0 + \hat{\phi}$, the equations
for the mean value and the fluctuations of the field read \cite{jppfdm}
\vspace{-5mm}
\begin{eqnarray}
\label{eq1} \Box \phi_0 + (m_B^2+\xi_B R) \phi_0 + \frac{\lambda_B}{3!} \phi_0^3 + \frac{\lambda_R}{2} \phi_0 \langle \hat{\phi}^2 \rangle = 0
\\ \Box \hat{\phi} + \bigg( m^2_R+\xi_R R + \frac{\lambda_R}{2} \phi_0^2 \bigg) \hat{\phi} = 0 \, .
\end{eqnarray}
The quantity $ \langle \hat{\phi}^2 \rangle$ is divergent and the infinities must be absorbed into the bare constants. 
Note that, as $ \langle \hat{\phi}^2 \rangle$ is already $O(\hbar)$, we replaced the bare constants by the renormalized ones
in the equation for the fluctuations and in the last term of the mean value equation.

As for the EMT, in order to absorb the infinities into bare constants we will subtract the adiabatic expansion of $\langle\hat\phi^2\rangle$. We will insert the definition 
\vspace{-2mm}
\begin{equation}
\langle \hat{\phi}^2 \rangle_{ren}=\langle \hat{\phi}^2 \rangle-\langle \hat{\phi}^2 \rangle_{ad}
\end{equation}
\vspace{-2mm}
into the field equation
\begin{equation}
\Box \phi_0 + (m_B^2+\xi_B R) \phi_0 + \frac{\lambda_B}{3!} \phi_0^3 + \frac{\lambda_R}{2} \phi_0 \langle \hat{\phi}^2 \rangle_{ren}
+\frac{\lambda_R}{2} \phi_0 \langle \hat{\phi}^2 \rangle_{ad} = 0\, ,
\end{equation}
and will analyze the divergences of $\langle\hat\phi^2\rangle_{ad}$. For the same reasons as before, we choose the adiabatic expansion
to perform the subtraction and renormalize $\langle\hat\phi^2\rangle$. 
A complete analysis using a covariant approach
and for general metrics can be found in Ref.\cite{jppfdm}. Here we will restrict ourselves to the case of FLRW metrics, we will consider a 
time-dependent mean value $\phi_0(t)$ and will regularize the theory 
using UV cutoffs. 

The fluctuation field $\hat\phi$ is a free field with a variable mass $M^2=m_R^2+\lambda_R\phi_0^2/2$.  As we are assuming that the mean value 
depends only on time, we can describe the field in terms its  Fourier modes, that satisfy Eq.\eqref{modes} with $\omega_k^2=k^2+M^2 a^2$.
 From Eq.\eqref{WKB} we have
\begin{equation}
\label{phi2}
\langle  \hat{\phi}^2  \rangle = \frac{1}{4 \pi^2 a^2} \int dk\,\frac{k^2}{W_k}\, .
\end{equation}
The function $W_k$ satisfies Eq.\eqref{nonl} and can be solved using the adiabatic approximation. Although now not only the scale factor but also $\phi_0$
depends on time, in principle one should include in the adiabatic expansion terms with derivatives of $\phi_0$. However, it can be shown that 
these terms give finite contributions to $\langle\hat\phi^2\rangle$, and therefore they can be omitted when discussing the renormalization
of the mean value equation in the one loop approximation \cite{jppfdm} (there is no wave function renormalization for $\lambda\phi^4$ theory in the 
one loop approximation).

Inserting the adiabatic approximation for $W_k$ into Eq.\eqref{phi2} we obtain \cite{Bunch}
\begin{equation}
\label{phi2addiv}
\langle \hat{\phi}^2 \rangle_{ad}^{div}= \frac{1}{4\pi^2a^2} \int dk \, k^2\bigg[\frac{1}{\omega_k} - \frac{\big(\xi_R-\frac{1}{6}\big)Ra^2}{2\omega_k^3} \bigg]\, .
\end{equation}

In order to see explicitly the differences between the different regularization methods, we quote the result obtained within dimensional regularization.
Replacing $k^2\to k^{n-2}$ in the integrand of Eq.\eqref{phi2addiv} and performing the integrations we obtain
\begin{equation}
\label{phi2n}
\langle\hat\phi^2\rangle_{ad}^{div} = \frac{1}{8\pi^2(n-4)} \bigg[m_R^2+\frac{\lambda_R}{2} \phi_0^2+\bigg(\xi_R-\frac{1}{6}\bigg)R\bigg]\, ,
\end{equation}
where we omitted finite contributions. 
The first term renormalizes the bare mass $m_B$, the second term the coupling constant $\lambda_B$, and the third one the coupling to the curvature $\xi_B$.

On the other hand, when using an UV cutoff $k_{UV}$
\begin{equation}
\langle\hat\phi^2\rangle_{ad}^{div}=\frac{1}{8\pi^2}\left[\frac{k^2_{UV}}{a^2}-\bigg(m_R^2+\frac{\lambda_R}{2}\phi_0^2+\bigg(\xi_R-\frac{1}{6}\bigg)R \bigg)\log\bigg(\frac{k_{UV}}{a \mu}\bigg)\right]\, ,
\label{phi2UV}
\end{equation}
where $\mu$ is an arbitrary scale and we omitted finite terms. 

Let us first consider a physical cutoff $k_{UV}=a \Lambda_{UV}$. From Eq.\eqref{phi2UV} we see that the quadratic divergence renormalizes the mass, while the logarithmic divergence
gives an additional term to the renormalization of the mass and renormalizes the other bare constants of the theory. The counterterms are, explicitly:
\begin{eqnarray}
m_R^2 &=& m_B^2 + \frac{\lambda_R}{16 \pi^2} \bigg[ \Lambda_{UV}^2 - m_R^2 \log (\Lambda_{UV}/\mu) \bigg] 
\\ \xi_R &=& \xi_B - \frac{\lambda_R \big(\xi_R-\frac{1}{6}\big)}{16 \pi^2} \log (\Lambda_{UV}/\mu) 
\\ \lambda_R &=& \lambda_B - \frac{3}{16} \lambda_R^2 \log (\Lambda_{UV}/\mu)\, .
\end{eqnarray}
There are some similarities with the situation in the renormalization of the SEE. With an UV cutoff, the bare mass contains a quadratic divergence
that is independent of the mass of the field. The logarithmic divergences has the same structure than in dimensional regularization. 

On the other hand, for a comoving cutoff we set $k_{UV}=\Lambda_c$ in Eq.\eqref{phi2UV}. We see that this choice presents an additional complication. The quadratic divergence
depends on the scale factor, and therefore cannot be absorbed into a redefinition of the mass \cite{foot3}. It is necessary to introduce an additional, non covariant term
in the classical action of the interacting field
\begin{equation}\label{actionwithsigma}
S_{matter}= \int d^4x \sqrt{-g} \bigg(\frac{1}{2} \phi_{,\mu} \phi^{,\mu} - \frac{1}{2}(m_B^2+\xi_B R) \phi^2 - \frac{\sigma_B}{a^2}\phi^2 - \frac{\lambda}{4!} \phi^4\bigg)\, .
\end{equation}

Assuming that the renormalized value of the new bare constant $\sigma_B$ is zero, that is, choosing
\vspace{-1mm}
\begin{equation}
 \sigma_R =0= \sigma_B + \frac{\lambda_R \Lambda_c^2}{16 \pi^2} \, , 
 \end{equation}
we obtain
\begin{eqnarray}
m_R^2 &=& m_B^2 - \frac{\lambda_R m_R^2}{16 \pi^2} \log (\Lambda_{c}/\mu) 
\\ \xi_R &=& \xi_B - \frac{\lambda_R \big(\xi_R-\frac{1}{6}\big)}{16 \pi^2} \log (\Lambda_{c}/\mu) 
\\ \lambda_R &=& \lambda_B - \frac{3}{16} \lambda_R^2 \log (\Lambda_{c}/\mu) 
\end{eqnarray}
which are similar to the counterterms obtained within dimensional regularization.

It is interesting to remark that, as in the case of the SEE, the logarithmic divergences have always the same structure. As a consequence, the $\beta$-functions of the theory will not depend on the regularization. On the contrary, the renormalized equation for the mean value (and therefore the effective potential for the scalar field) does depend on the 
regularization. Indeed, the analysis can be perfomed along the lines of Section IIIC, to check explicitly that the mean value equations in different regularizations will differ
by local terms.

It is also worth to note that the presence of the  term proportional to $\sigma_B/a^2$ in Eq.(\ref{actionwithsigma}) introduce an ambiguity in the SEE, and could complicate its renormalization. Indeed, in order to obtain the SEE in FLRW spacetimes, one should write the action in terms of the scale factor and the lapse function $N$, and the extra factor
$1/a^2$ can be written in different ways in terms of $a$ and $N$.

\section{Conclusions}\label{sec:concl}

In this paper we have analyzed the renormalization process for a scalar field FLRW sacetimes, using UV cutoffs. Our main findings are the following:
\begin{itemize}
\item If one assumes that the infinities of the theory should be absorbed into bare constants of the theory, the subtraction is crucial and involves the adiabatic EMT.
\item The divergences that result in the case of a comoving cutoff are conserved tensors, and can be absorbed into the gravitational constants and additional bare fluids that should be introduced ad-hoc into the theory.
\item For a physical cutoff, the adiabatic EMT is not conserved, and it is necessary to introduce non covariant counterterms. Fine tuning of the finite parts 
of the counterterms  is needed to restore covariance at the renormalized level.
\item All the regularizations give different answers for the effective EMT that appears in the SEE, even using the same subtraction. In particular, both the comoving and physical cutoffs do not produce an anomalous trace in the effective EMT.
\item The logarithmic divergences are always proportional to geometric tensors.
\item In contrast with what happens in the SEE for free fields, when considering the mean value equation for interacting fields, the use of a comoving cutoff needs the introduction of
non covariant counterterms to renormalize the theory. 
\end{itemize}

Regarding previous works on the subject, we have showed explicitly that the renormalization suggested in Refs.\cite{Maggiore2012,Maroto1} can indeed be pursued. However,
we have described important aspects that have been overlooked before:  on the one hand, adiabatic subtraction is crucial to absorb the infinities into the  bare constants
of the theory. One cannot make a subtraction based on each particular
metric considered \cite{Maroto1,Maroto2}.  Moreover,  even after renormalization, the final SEE are different than those coming from the usual approach, 
and different for comoving and physical
cutoffs. 
On the other hand, there is no reason to take the bare value of the EMT as its physical value, even after discarding the quartic divergences,  as done in Refs.\cite{Bilic,Bernard}. 

In summary, the use of UV cutoffs, although at first sight more physical than dimensional regularization, presents additional  complications and 
arbitrariness, and in our opinion do not shed additional light on the cosmological constant and dark energy problems. 

We would like to end the paper with a comment related to the use of infrared  (IR) cutoffs. 
Present knowledge establishes that the seeds for the inhomogeneities
  in the universe are quantum fluctuations that become classical when their  wavelengths become larger than the horizon \cite{booksinf}. It is therefore natural to consider an 
  effective field theory for the long wavelengths modes of the field, integrating out the short wavelengths. Therefore, the quantum part of the field is restricted
  to  modes with $k>H a(t)$ and one has to face the problem of renormalizing the UV divergences in the presence of a time dependent IR cutoff. This is of course not
  restricted to cosmological applications.  The same issue is present in the study of phase transitions, where the long wavelength part of the field associated to the
  order parameter of the transition becomes classical and non homogeneous, giving rise to the formation of domains \cite{rivers}.  A naive generalization of the usual adiabatic
  renormalization to the case of a time dependent IR cutoff gives of course a non-conserved energy-momentum tensor, even for free fields and using 
   a covariant regularization. The non conservation is in this case physical,
  since during time evolution new modes enter into the classical field, and therefore there is energy exchange between the quantum and classical parts of the field.
  Only the full EMT, including the quantum, classical and stochastic parts should be conserved.
  Alternatively, one could  consider comoving IR  cutoffs \cite{diana}. In both cases, dimensional regularization can be used to deal with the UV divergences.
\vspace{-10mm}
\appendix
\section{} \label{appendix1}
\vspace{-5mm}
In this appendix we collect some useful formulas used in the calculations. 
\vspace{-2mm}

The explicit expressions for the geometric tensors in FLRW metrics are
\vspace{-3mm}
\begin{eqnarray}\label{A1}
R &=&\frac{3}{a^2} \bigg(D'+\frac{1}{2}D^2 \bigg) \nonumber \\ [2mm]
G_{00} &=& -\frac{3}{4} D^2 \nonumber \\[2mm]
G_{ii} &=& D'+\frac{D^2}{4} \nonumber \\[3mm]
H_{00}^{(1)} &=& \frac{9}{a^2} \bigg(\frac{1}{2}D'^2 - D''D +\frac{3}{8} D^4 \bigg) \nonumber \\[3mm]
H_{ii}^{(1)} &=& \frac{3}{a^2} \bigg( 2 D'''- D''D+\frac{1}{2}D'^2 - 3D'D^2+\frac{3}{8}D^4 \bigg),
\end{eqnarray}
where $D=2a'/a$.

The convergent part of the adiabatic stress tensor mentioned in Eq.(\ref{Tdivcon}) is given by \cite{Bunch}
\begin{eqnarray}
\label{finite25}
\langle T_{00} \rangle_{ad}^{conv} &=& \frac{1}{8\pi^2a^2} \int dk k^2 \bigg[ \frac{a^4m^4D^2}{16 \omega^5}- \frac{a^4m^4}{64 \omega^7} (2D''D-D'^2+4D'D^2+D^4)\nonumber \\
&+& \frac{7a^6m^6}{64 \omega^9} (D'D^2+D^4) - \frac{105 a^8m^8D^4}{1024 \omega^{11}}+ \bigg(\xi-\frac{1}{6} \bigg) \bigg( \frac{a^2m^2}{8\omega^5}(6D''D-3D'^2+6D'D^2) \nonumber \\
&-& \frac{a^4m^4}{64\omega^7} (120D'D^2+105D^4) + \frac{105a^6m^6D^4}{64\omega^9} \bigg) + \frac{\big(\xi-\frac{1}{6} \big)^2 a^2m^2}{8\omega^5} (54D'D^2+27D^4) \bigg] \nonumber \\
\langle T^\alpha_{\ \alpha} \rangle_{ad}^{conv} &=& \frac{1}{4\pi^2a^4} \int dk k^2 \bigg[ \frac{a^4 m^4}{8 \omega^5} (D'+D^2) - \frac{5a^6m^6D^2}{32 \omega^7} \nonumber \\
&-& \frac{a^4m^4}{32\omega^7}(D'''+4D''D+3D'^2+6D'D^2+D^4) \nonumber \\
&+& \frac{a^6m^6}{128\omega^9}(28D''D+21D'^2+126D'D^2+49D^4) - \frac{231a^8m^8}{256\omega^{11}}(D'D^2+D^4) \nonumber \\
&+& \frac{1155a^{10}m^{10}D^4}{2048\omega^{13}} +  \bigg(\xi-\frac{1}{6} \bigg) \bigg( \frac{9a^4m^4D^2}{4\omega^5} + \frac{a^2m^2}{4\omega^5}(3D'''+6D''D+\frac{9}{2}D'^2+3D'D^2) \nonumber \\
&-& \frac{a^4m^2}{32\omega^7} (120D''D+90D'^2+390D'D^2+105D^4) + \frac{a^6m^6}{128\omega^9}(1680D'D^2+1365D^4) \nonumber \\
&-& \frac{945 a^8 m^8 D^4}{128 \omega^{11}} \bigg) +  \bigg(\xi-\frac{1}{6} \bigg)^2 \bigg( \frac{a^2m^2}{32\omega^5} (432D''D+324D'^2+648D'D^2+27D^4) \nonumber \\
&-& \frac{a^4m^4}{16\omega^7}(270D'D^2+135D^4) \bigg) \bigg]\, .
\end{eqnarray}

\section*{Acknowledgements}
FDM would like to thank D. L\'opez Nacir and L. Trombetta for interesting discussions on related matters, and ICTP for hospitality, where part of this work has been done.
This research was supported by ANPCyT, CONICET, and UNCuyo. 


\begin{thebibliography}{bib}
\bibitem{foot1}This is not the case if one computes the function $W_k$ in four dimensions and regularizes only
the integration measure in momentum space.
\bibitem{foot2}Note that this does not mean loss of generality, since we are adding and subtracting $\langle T_{\mu\nu}\rangle_{sub}$  in the SEE (see Eq.(\ref{sub}).
\bibitem{foot4} As already stressed, for a physical cutoff there are no contributions proportional to $\log(a)$, that were crucial for the EMT conservation in Eq.(\ref{extracom}).
\bibitem{foot3} This point has been noticed in Ref.\cite{Maroto2}.
\end{thebibliography}
\end{document}